\documentclass{ws-procs975x65}

\newcommand{\nno}{\nonumber}

\newcommand{\isdef}{\mathrel{\mathop:}=}

\newcommand{\la}{\langle}
\newcommand{\ra}{\rangle}

\newcommand{\be}{\begin{equation}}
\newcommand{\ee}{\end{equation}}
\newcommand{\bea}{\begin{eqnarray}}
\newcommand{\eea}{\end{eqnarray}}

\newcommand{\hf}[4]{F\!\left[#1,#2;#3;#4\right]}

\newcommand{\psf}[1]{\psi\!\left(\!#1\!\right)}

\newcommand{\psr}{\la\Phi^{2}\ra_{\text{ren}}}
\newcommand{\tmn}{\la T_{\mu\nu}(x)\ra_{\text{ren}}}

\begin{document}

\title{SCALAR FIELD HADAMARD RENORMALISATION IN $AdS_{n}$}
\author{CARL KENT$^*$ and ELIZABETH WINSTANLEY$^{**}$}
\address{School of Mathematics and Statistics, University of Sheffield,\\
Sheffield S3 7RH, U.K.\\
$^*$e-mail: app09ck@sheffield.ac.uk, $^{**}$e-mail: e.winstanley@sheffield.ac.uk}
\begin{abstract}
We outline an analytic method for computing the renormalised vacuum expectation value of the quadratic fluctuations and stress-energy tensor associated with a quantised scalar field propagating on $AdS_{n}$.  Explicit results have been obtained using Hadamard renormalisation in the case of a massive neutral scalar field with arbitrary coupling to the curvature, for $n=2$ to $n=11$ inclusive.
\end{abstract}
\bodymatter
\section{Introduction}
Hadamard renormalisation (HR) is a rigorous and elegant approach to computing expectation values in quantum field theories in curved space.  The defining feature of HR is that it exploits the singularity structure of Hadamard's `elementary solution'\,\cite{Had} of second order 
differential equations.  HR has proven to be a powerful technique for four-dimensional space-times, but so far it has been put to little use in  
higher dimensional space-times.

Following D\'{e}canini and Folacci's scheme for HR in a general space-time\,\cite{D&F08}{, we} find the renormalised vacuum expectation value (v.e.v.) of the quadratic fluctuations and stress-energy tensor associated with a quantised neutral scalar field in $AdS_{n}$ with a general coupling $\xi$. Detailed accounts of this work are to appear in Ref.~\refcite{kw13}.
\section{Scalar field propagation on $AdS_{n}$}
The maximal symmetry of $AdS_{n}$ means two-point scalar functions $f(x,x')$ defined on it only depend on the distance $s(x,x')$ separating the points along a shared geodesic.  Information about the propagation of scalar fields having quanta of mass $m$ and coupled with strength $\xi$ to a scalar curvature $\mathcal{R}$ is encapsulated in the scalar field propagator.  When represented by the Feynman Green function $G_{F}(x,x')$ (divergent as $x'\to x$), this propagator is a solution to the inhomogeneous scalar field wave equation, which for a maximally symmetric space may be written as
\be
\left(\Box-m^{2}-\xi\mathcal{R}\right)G_{F}(\sigma)=g^{-\frac{1}{2}}\delta(\sigma)\quad\mid\quad\sigma\isdef\frac{1}{2}s^{2}\quad,\quad g\isdef|\!\det g_{\mu\nu}|,
\label{kw:eq2}
\ee
where accordingly, $G_{F}(\sigma)$ diverges as $\sigma\to0$.
\section{Hadamard form of the propagator $G_{F}^{H}(\sigma)$}
Hadamard's `elementary solution'\,\cite{Had} means the propagator has the form
\be
G_{F}^{H}(\sigma)=i\nu(n)\left[U(\sigma)\sigma^{1-\frac{n}{2}}+V(\sigma)\ln\bar{\sigma}+W(\sigma)\right]\quad\mid\quad V(\sigma)=0~~\forall~n\text{ odd},
\label{kw:eq4}
\ee
where $\nu(n)$ is a constant and $U(\sigma)$, $V(\sigma)$ and $W(\sigma)$ are regular functions as $\sigma\to0$.

The superscript $H$ in \eref{kw:eq4} denotes the Hadamard form, distinguishing it from expressions derived independently from the dynamics of the theory in \eref{kw:eq2}. The Hadamard form splits into a purely geometric part
\be
G_{F\text{, sing}}^{H}(\sigma)\isdef i\nu(n)\left[U(\sigma)\sigma^{1-\frac{n}{2}}+V(\sigma)\ln\bar{\sigma}\right]\quad\mid\quad V(\sigma)=0~~\forall~n\text{ odd},
\label{kw:eq20}
\ee
(the source of the propagator's divergence), and a regular state-dependent part
\be
G_{F\text{, reg}}^{H}(\sigma)\isdef i\nu(n)W(\sigma).
\ee
When $n$ is even, $G_{F\text{, sing}}^{H}(\sigma)$ contains, as $\sigma\to0$, a non-vanishing finite term $f_{0}$.  In \eref{kw:eq20} we define $\bar{\sigma}\isdef m_{*}^{2}\sigma$, where $m_{*}$ is a mass renormalisation scale that is introduced to ensure a dimensionless logarithmic argument.  Similarly, we define $\bar{a}\isdef m_{*}a$, where $a$ is the radius of curvature of $AdS_{n}$.
\section{Renormalised v.e.v. of the quadratic field fluctuations $\psr$}
Allen and Jacobson \cite{A&J} have derived an expression for the propagator from \eref{kw:eq2} involving a linear combination of hypergeometric functions ${F\isdef{}_{2}F_{1}}$, with constant coefficients denoted by $C$ and $D$:
\be
G_{F}(\sigma)\!=\!C\hf{\frac{n-\!1}{2}\!+\!\mu}{\frac{n-\!1}{2}-\mu}{\frac{n}{2}}{z}\!+\!D\hf{\frac{n-\!1}{2}\!+\!\mu}{\frac{n-\!1}{2}-\mu}{\frac{n}{2}}{1\!-\!z}\!\!.
\label{kw:eq5}
\ee
The hypergeometric functions in \eref{kw:eq5} have arguments depending on {${z=z(\sigma)}$} (where $z\to1$ as $\sigma\to0$), and orders depending on the constant
\be
\mu\isdef\sqrt{\frac{\left(n-1\right)^{2}}{4}+m^{2}a^{2}+\xi\mathcal{R} a^{2}}.
\ee
The pertinent point of the form in \eref{kw:eq5} is that the `$C$-term' is regular and that the `$D$-term' is singular as $\sigma\to0$.

We compute $\psr$ using HR as follows:
\be
\psr\isdef-i\lim_{\sigma\to0}\left[G_{F}(\sigma)-G_{F\text{, sing}}^{H}(\sigma)\right],
\label{kw:eq6}
\ee
so that for a finite $\psr$, singular terms in $G_{F}(\sigma)$ must equal those in $G_{F\text{, sing}}^{H}(\sigma)$.  Note that for even $n$, \eref{kw:eq6} includes the subtraction of the finite term $f_{0}$.

As examples, the explicit results for $n=6$ and $n=7$ respectively are
\be
\hspace{-2.1mm}\psr\!=\!-\frac{1}{64\pi^{3}a^{4}}\!\left\{\!\!\left(\!\mu^{4}-\frac{5}{2}\mu^{2}+\frac{9}{16}\right)\!\!\left[\psf{\frac{1}{2}+\mu}\!-\ln\bar{a}\right]\!-\frac{3}{4}\mu^{4}+\frac{29}{24}\mu^{2}+\frac{107}{960}\!\right\}\!,
\ee
\be
\psr=-\frac{1}{240\pi^{3}a^{5}}\left\{\mu^{5}-5\mu^{3}+4\mu\right\}.
\ee
Expressions for $\psr$ consist of even or odd powered polynomials in $\mu$ of leading order $n-2$ that are constant everywhere in space-time.  Expressions for even $n$ include an additional factor containing a psi function and a logarithmic term.
\section{Renormalised v.e.v. of the stress-energy tensor $\tmn$}
The computation of $\tmn$ follows essentially the same method as that of $\psr$.  However, the definition of $\tmn$ involves the action of a second-order linear differential operator $\mathcal{T}_{\mu\nu}(x,x')$ on $G_{F}(\sigma)$.  For even $n$, there is an additional non-vanishing locally conserved tensor \cite{D&F08} $\Theta_{\mu\nu}(x)$.
\be
\tmn=-i\lim_{x'\to x}\mathcal{T}_{\mu\nu}(x,x')\left[G_{F}(\sigma)-G^{H}_{F,\text{ sing}}(\sigma)\right]+\Theta_{\mu\nu}(x).
\label{kw:eq9}
\ee
Explicit results are given below for $n=4$ and $n=5$ respectively.  The detailed form of $\Theta_{\mu\nu}(x)$ for $n=4$ is omitted here.  It can be found in Ref.~\refcite{D&F08}.
\bea
\tmn=&&\!\!\frac{3}{128\pi^{2}a^{4}}\!\left\{\!\left(-\frac{4}{3}\mu^{4}-\!\left(\!16\xi-\frac{10}{3}\right)\!\mu^{2}+4\xi-\frac{3}{4}\right)\!\!\left[\psf{\frac{1}{2}+\mu}\!-\ln\bar{a}\right]\right.\nno\\
&&\quad\qquad\quad\left.+\,\mu^{4}+\!\left(\!8\xi-\frac{29}{18}\right)\!\mu^{2}+\frac{2}{3}\xi-\frac{107}{720}\right\}\!g_{\mu\nu}(x)+\Theta_{\mu\nu}(x),
\eea
\be
\tmn=\frac{1}{120\pi^{2}a^{5}}\!\left\{\mu^{5}+\left(20\xi-5\right)\!\mu^{3}-\left(20\xi-4\right)\!\mu\right\}\!g_{\mu\nu}(x).
\ee
Expressions for $\tmn$ are proportional to the metric $g_{\mu\nu}(x)$ of $AdS_{n}$ with constants of proportionality consisting of even or odd powered polynomials in $\mu$ of leading order $n$ and linearly dependent on $\xi$.
\section{Summary and outlook}
Expressions have been obtained for $\psr$ and $\tmn$ explicitly for {$n=2$} to $n=11$ inclusive using HR and agree with results obtained using $\zeta$-function regularisation\,\cite{Cald}{. With} sufficient computational power and time, expressions could be generated for any $n$.

In addition to this study of the vacuum state of quantised scalar fields coupled to $AdS_{n}$, work is currently in progress on rotating and thermal scalar field states\,\cite{kw13}\!\! .
\section*{Acknowledgments}
This research is supported by a UK EPSRC DTA studentship, the Lancaster-Manchester-Sheffield Consortium for Fundamental Physics under grant ST/J000418/1 and European Cooperation in Science and Technology (COST) action MP0905 ``Black Holes in a Violent Universe''.
\bibliographystyle{ws-procs975x65}
\bibliography{main}
\end{document}